\documentclass[prd,10pt, tightenlines, twocolumn]{revtex4-1}
\input epsf.tex
\epsfclipon

\usepackage{slashed}
\usepackage{color}

\usepackage{epsfig}
\usepackage{epstopdf}
\usepackage{epsf}
\input{epsf.sty}
\usepackage{graphicx,amsmath,amssymb}
\usepackage{multirow}

\allowdisplaybreaks

\usepackage{bbm,bm,amsmath,amssymb}
 
\usepackage[normalem]{ulem}

\usepackage{comment}
\def\be{\begin{equation}}
\def\ee{\end{equation}}
\def\figs/B{B}
\def\bea{\begin{eqnarray}}
\def\eea{\end{eqnarray}}
\def\bg{\begin{eqnarray}}
\def\nd{\end{eqnarray}}

\def\log{{\rm log}}

\begin{document}

\title{Chiral Symmetry and the Cosmological Constant} 
\author{Stephon Alexander, Gabriel Herczeg, Jinglong Liu, and Evan McDonough}
\affiliation{ Brown Theoretical Physics Center and Department of Physics, Brown University, 182 Hope Street, Providence, Rhode Island, 02903}


\begin{abstract}
In this work we provide a link between a nearly vanishing cosmological constant and chiral symmetry.  This is accomplished with a modification of General Relativity coupled to a topological field theory, namely BF theory, by introducing fermions charged under the BF theory gauge group. We find that the cosmological constant sources a chiral anomaly for the fermions, providing a `technical naturalness' explanation for the smallness of the observed cosmological constant.  Applied to the early universe, we show that production of fermions during inflation can provide all the dark matter in the universe today, in the form of superheavy dark baryons. \end{abstract}

\maketitle

\section{Introduction}

The cosmological constant reigns as one of the oldest and most challenging puzzles in contemporary physics. There is vast literature defining and debating cosmological constant problems \cite{Zeldovich:1967gd,weinberg1989cosmological, padilla2015lectures, Bousso2007TheCC}, and a long history of attempts at solutions. 
Parallel to this, a central problem in modern cosmology is to understand the nature of dark energy; the mysterious energy source driving the accelerated expansion of the universe. Evidence for dark energy was first observed in the relationship between the luminosity and redshift of distant supernovae \cite{perlmutter1999measurements, riess1998observational}, and subsequently confirmed by observations of the cosmic microwave background and large scale structure \cite{2013PhR...530...87W}. The most recent analysis of the Planck experiment  \cite{aghanim2018planck} gives  the fraction of the universe comprised of dark energy as $\Omega_\Lambda = 0.6847 \pm 0.0073$, differing from $0$ at $>93\sigma$.

Perhaps the simplest explanation for dark energy is that the evolution of our universe is described by general relativity with a small, positive cosmological constant. This demands a serious investigation of the cosmological constant problem, which itself long outdates the observations of dark energy (e.g.~\cite{weinberg1989cosmological} and references therein). Motivated by this, the starting point for this work is to attempt to understand why quantum contributions to the cosmological constant do not gravitate.

One approach to the cosmological constant problem is to directly modify the Einstein-Hilbert action, in a manner so as to protect the cosmological constant from loop corrections and renormalization. This was discussed in detail already in 1989 \cite{weinberg1989cosmological}. These include modifying the action to include a volume averaging \cite{Tseytlin:1990hn,Gabadadze:2014rwa,Gabadadze:2016avp}, the realization of this in a 5d spacetime \cite{Gabadadze:2015goa}, and an  effective `sequestering' of the cosmological constant \cite{Kaloper:2013zca,Kaloper:2014dqa,Kaloper:2016jsd} through the use of Lagrange multipliers. All of these directly modify the Einstein-Hilbert action to resolve aspects of the cosmological constant problem.

A recent proposal along these lines \cite{Alexander:2018tyf} is to couple General Relativity to a topological field theory, namely BF theory, with the volume form of the Einstein-Hilbert action replaced by a wedge product of B-fields. In this scenario, loop corrections to the vacuum energy imbue dynamics on the $B$ and $F$ fields, without affecting the curvature of spacetime. In this sense, the quantum aspects of the cosmological constant problem are not present in this model, at least at the level of the action (though not necessarily in the path integral). This is done without promoting the cosmological constant to a dynamical field,  and relies on spacetime-dependent fields in the BF sector. Both of these constitute an evasion of the classic no-go theorem \cite{weinberg1989cosmological}.

A simple generalization of \cite{Alexander:2018tyf}, which retains its essential properties (in particular, the evasion of no-go theorems), is that to a matter-coupled topological field theory, through the introduction of fermions or complex scalars charged under the BF-theory gauge group. Such theories, particularly matter-coupled Chern-Simons theory, have been extensively studied, notably in the context of ABJM theory \cite{Aharony:2008ug}; see also \cite{Gates:1991qn,Nishino:1991sr,Nishino:1996xb} for early work in this direction. Coupling to matter generically breaks the topological nature of theory, as it admits local excitations, but the topological action emerges in the IR as the lowest-order terms in  the derivative expansion. In the context of \cite{Alexander:2018tyf}, the topological nature of BF theory is already spoiled by coupling to gravity, and thus it is natural to consider incorporating matter charged under the gauge group of the BF theory.  In this paper we consider the physics of Dirac fermions charged under the BF gauge group of \cite{Alexander:2018tyf}, and their impact on the cosmological constant.

A striking implication of the matter-coupled theory is a chiral anomaly of the charged fermions, generated by the $\mathbf{F}$ field of the BF-theory, that is sourced by the cosmological constant. In the limit $\Lambda \rightarrow 0$, the $U(1)$ axial symmetry is restored, giving a \emph{technical naturalness} \cite{tHooft:1979rat} argument for the smallness of the observed the cosmological constant.  This connects the physics of the deep IR, namely the curvature of spacetime, to the physics of the UV, namely vacuum production of particles, and, in so doing, provides a quantum solution to a seemingly classical aspect of the CC problem.

An implication of the chiral anomaly is that any non-zero cosmological constant leads to the vacuum production of fermions.  We find that the observed value of $\Lambda$ gives a negligibly small amount of particle production, and thus does not the standard cosmological evolution. However we find that an early phase of accelerated expansion, e.g. cosmic inflation \cite{Guth:1980zm}, can induce significant particle production, and in fact, provide the correct relic density of dark matter as we observe universe today. A simple scenario for dark matter that can arise in this way is a model of dark baryons, see e.g.~\cite{Antipin:2015xia,Hertzberg:2019bvt}.

The structure of this paper is as follows: In section \ref{BFgravity} we review BF-coupled gravity, and the non-renormalization of the cosmological constant. In Section \ref{sec:mattercoupled} we introduce the matter-coupled theory and its main features, and in Section \ref{DEasCA} construct a solution exhibiting a CC-induced chiral anomaly, as mentioned above. In Section \ref{sec:presentday} we compute the chiral asymmetry generated by the observed cosmological constant and find it is well within observational bounds, while in \ref{sec:inflation} we consider the chiral asymmetry generated by inflation, and find it can explain the observed dark matter abundance. We conclude in section \ref{sec:conclusion} with a discussion of directions for future work.

\section{Non-Renormalization of the cosmological constant in BF-Coupled Gravity}

\label{BFgravity}

In this section, we give an overview of a model proposed in \cite{Alexander:2018tyf} that solves aspects of the cosmological constant problem. The approach taken in that article couples a conformalized version of Einstein gravity to a topological field theory, namely BF theory. We review the key features of BF theory, and introduce the model presented in \cite{Alexander:2018tyf} as our starting point, giving particular attention to the fact that the cosmological constant that gravitates in that model does not receive corrections from quantum loops, thus solving important aspects of the cosmological constant problem.

\subsection{BF theory}
We now give a very brief overview of BF theory, giving particular attention to the classical field-theoretic aspects. For a thorough introduction to the subject including the quantum theory and relations to computing sopolical invariants, see for example \cite{cattaneo1995topological}. For the extension to \emph{supersymmetric} BF-theory, which is the case most relevant to string theory, see, e.g., \cite{Brooks:1994nn}.

 Let $M$ be a smooth manifold, $G$ be a semi-simple Lie group, and let $\pi:P\to M$ be a $G$-principal bundle over $M$, equipped with a connection $\mathbf{A}$ which we view as a $\mathfrak{g}$-valued one-form on $M$, where $\mathfrak{g}$ is the Lie algebra of $G$. In any number of spacetime dimensions $n = dim(M)$, one can define a topological field theory via the so-called BF action:
\be
S_{BF} = \int_M \mbox{tr}(\mathbf{B}\wedge \mathbf{F}),
\ee
where $\mathbf{B}$ is an adjoint valued $(n-2)$-form, and $\mathbf{F} = d\mathbf{A} + \mathbf{A} \wedge \mathbf{A}$ is the curvature two-form of the connection $\mathbf{A}$. We consider an arbitrary semi-simple gauge group, in which case the trace appearing in the action is understood to be defined in terms of the non-degenerate Killing form associated on the Lie algebra.

In three or four dimensions, it is possible to include an additional term in the action depending on a ``cosmological constant" $\mu$. In three dimensions, one can consider
\be
S_3 = \int_M \mbox{tr}(\mathbf{B}\wedge \mathbf{F} + \frac{\mu^2}{3}\mathbf{B} \wedge \mathbf{B} \wedge \mathbf{B}).
\ee
When the gauge group $G = SO(2,1)$ is chosen, it can be shown that $I_{3}$ is a first-order action for (2+1)-dimensional Einstein gravity formulated in terms of a vierbien and spin connection. In  four dimensions, one can consider
\be
S_4 = \int_M \mbox{tr}(\mathbf{B}\wedge \mathbf{F} + \frac{\mu}{2}\mathbf{B}\wedge \mathbf{B}).
\ee
In what follows, we will restrict our considerations to four spacetime dimensions, so let us concentrate our focus on $S_4$. Varying $S_4$ with respect to $\mathbf{A}$ and $\mathbf{B}$ respectively give
\be
d_{\mathbf{A}}\mathbf{B} = 0, \qquad \mathbf{F} + \mu\mathbf{B} = 0,
\ee
where $d_{\mathbf{A}}\mathbf{Q} = d\mathbf{Q} + [\mathbf{A}, \mathbf{Q}]$. 

The action $S_4$ is invariant under two sets of gauge transformations. First, we have invariance under the usual transformation of the connection $\mathbf{A}$, provided $\mathbf{B}$ transforms accordingly. Given a local gauge parameter $g$, $S_4$ is invariant under the combination
\be
\mathbf{A} \to g^{-1}\mathbf{A}g + g^{-1}dg, \qquad \mathbf{B} \to g^{-1}\mathbf{B}g.
 \label{firstGauge}
\ee
Second, $S_4$ is invariant under transformations parametrized by a Lie algebra valued one-form $\boldsymbol{\eta}$ given by
\be
\mathbf{A} \to \mathbf{A} + \mu\boldsymbol{\eta}, \qquad  \mathbf{B} \to \mathbf{B} - d_{\mathbf{A}}\boldsymbol{\eta}. \label{secondGauge}
\ee
The latter set of gauge transformations imply that $\mathbf{B}$ is locally pure gauge, thus only the topological features of $\mathbf{B}$ are relevant. In the gravity model of \cite{Alexander:2018tyf} which we are now ready to introduce, this will no longer be the case: the action we will study is invariant under the usual gauge transformations \eqref{firstGauge} but not the second set of transformations \eqref{secondGauge}. This is a good thing, since in what follows we interpret $\textrm{tr}(\mathbf{B}\wedge \mathbf{B})$ as a spacetime volume form, but a transformation of the form \eqref{secondGauge} can always be used to set $\textrm{tr}(\mathbf{B}\wedge \mathbf{B})$ to zero. 

\subsection{BF-Coupled Gravity}

The theory we propose has the remarkable feature that it contains the Einstein field equation where the cosmological constant drops out and reappears in a hidden BF sector without disrupting the visible sector.  We begin by considering a theory coupling BF theory \cite{10.1007/3-540-46552-9_2,doi:10.1063/1.531238} with gravity, as proposed in \cite{Alexander:2018tyf}. The action reads
\begin{equation}
S=\int_M {\rm tr}(\mathbf{B}\wedge \mathbf{F})+\biggl[\frac{1}{2\kappa}R(\hat{g})+\frac{\bar{\Lambda}}{\kappa}+\mathcal{L}_M\biggr] {\rm tr}(\mathbf{B}\wedge\mathbf{B}). \label{action}
\end{equation}
where $\mathbf{F} = d{\mathbf{A}} + {\mathbf{A}}\wedge {\mathbf{A}}$ is the field strength two form associated with the non-abelian gauge field ${\mathbf{A}}$, $\mathbf{B}$ is a \emph{maximal rank}, adjoint-valued two-form, and we restrict attention to the case $dim(M) = 4$. The quantity $R(\hat{g})$ appearing in the action is the Ricci scalar associated with the Lorentzian metric $\hat{g}_{\mu\nu}$ which is defined as follows. Since we assume that $\mathbf{B}$ is maximal rank, ${\rm tr}(\mathbf{B}\wedge\mathbf{B})$ is a volume form on $M$. This volume form determines a density $\omega$, represented in a given coordinate system by 
\begin{equation}
{\rm tr}(\mathbf{B}\wedge\mathbf{B}) = \frac{\sqrt{\omega}}{4!}\epsilon_{\mu\nu\rho\sigma}dx^\mu\!\wedge dx^\nu\!\wedge dx^\rho\!\wedge dx^\sigma. \label{omegaDef}
\end{equation}
The action \eqref{action} depends on an arbitrary Lorentzian metric $g_{\mu\nu}$, while the ``physical" metric $\hat{g}_{\mu\nu}$ is defined in terms of $g_{\mu\nu}$ and $\mathbf{B}$ by
\begin{equation}
\hat{g}_{\mu\nu} = \left(\frac{\omega}{g}\right)^{\!1/4}\!g_{\mu\nu}.
\end{equation}
Note that \eqref{omegaDef} implies that $\sqrt{\hat{g}} = \sqrt{\omega}$, so that the information of the physical spacetime volume is completely encoded in the volume form $\mathbf{B}\wedge \mathbf{B}$, or equivalently, the density $\omega$.

Varying \eqref{action} with respect to $\mathbf{B}$, $\mathbf{A}$ and $g_{\mu\nu}$ respectively, leads to the following equations of motion
\begin{gather}
\mathbf{F}+\frac{1}{2\kappa}[4\bar\Lambda+R(\hat{g})+\kappa T]\mathbf{B}=0 \label{EOM1}\\
d_{\mathbf{A}} \mathbf{B}=0 \label{EOM2}\\
R_{\mu\nu}(\hat{g})-\frac{1}{4}R(\hat{g})\hat{g}_{\mu\nu}=\kappa\biggl(T_{\mu\nu}-\frac{1}{4}T\hat{g}_{\mu\nu}\biggr), \label{TFEE}
\end{gather}
where $T_{\mu\nu}$ is the stress tensor defined as the variation with respect to $\hat{g}$,
\begin{equation}
\label{defTab}
T_{\mu\nu}=-\frac{2}{\sqrt{|\omega|}}\frac{\delta\mathcal{S}_{\rm M}(\hat{g},\bm{\Phi})}{\delta\hat{g}^{\mu\nu}}=-2\frac{\partial\mathcal{L}_{\rm M}}{\partial \hat{g}^{\mu\nu}}+\mathcal{L}_{\rm M}\hat{g}_{\mu\nu}.
\end{equation}
The Einstein equation \eqref{TFEE} is \emph{traceless} Einstein equation. The trace of the Ricci curvature is determined by solving the three equations in conjunction.

This theory exhibits the interesting feature that $\bar\Lambda$, i.e. the quantity appearing in the action and path integral, is not the physical cosmological constant, i.e., it does not contribute to the curvature of spacetime. To see this, one can take the covariant exterior derivative of \eqref{EOM1}, and use equation \eqref{EOM2} to find,
\be
\label{dEOM}
{\rm d}(R(\hat{g})+\kappa T) = 0. 
\ee
From this, one can define an integration constant $C \equiv 4\Lambda$, to arrive at the equation of motion
\be
\label{RTLambda}
R(\hat{g})+\kappa T = 4 \Lambda . 
\vspace{0.1cm}
\ee
This, in conjunction with \eqref{TFEE}, fully specifies the behaviour of gravity and matter in the universe aside from the BF-theory. One can appreciate that the cosmological constant appearing is \emph{not} $\bar\Lambda$, which is subject to loop corrections, but instead $\Lambda$, which can be freely chosen.

We emphasize that this is not a \emph{cancelation}; the renormalized quantity $\bar\Lambda$ drops out of \eqref{dEOM} simply because it is, by definition, a constant. This implies that this procedure can protect the cosmological constant from renormalization at all times in the evolution of the universe, as opposed to a canonical renormalization group counter-term.

Thus this theory resolves one aspect of the cosmological constant problem: why is the cosmological constant not renormalized to a large value by vacuum bubbles in quantum field theory? However, another aspect remains: why is the observed cosmological constant \emph{small} by comparison to any other quantity in physics. This is sometimes referred to as the ``new" cosmological constant problem. In later sections, we will see this is naturally addressed by the inclusion of charged fermions.

\section{Extension to Matter-Coupled BF Theory}

\label{sec:mattercoupled}

Since the above model includes a dynamical gauge field $\bf{A}$, it is natural to couple the model also to a fermion field $\psi$, which for simplicity we take to be massless. For an arbitrary, semi-simple, non-abelian gauge group, such a massless fermion is described by the Dirac Lagrangian
\begin{equation}
\mathcal{L}_D= i\bar{\psi}\slashed{D}\psi = \bar{\psi}_i\gamma^\mu(\delta^i_{\,j}\,i\partial_{\mu} + gA^a_\mu(T^a)^i_{\, j})\psi^j. \label{1}
\end{equation}
where $g$ is the gauge coupling, and $T^a$ are generators of the gauge group. The full action becomes,
\begin{widetext}
\begin{equation}
S=\int_M {\rm tr}(\mathbf{B}\wedge \mathbf{F})+\biggl[\frac{1}{2\kappa}R(\hat{g})+\frac{\bar{\Lambda}}{\kappa} + \mathcal{L}_D + \mathcal{L}_M\biggr] {\rm tr}(\mathbf{B}\wedge\mathbf{B}), \label{full action}
\end{equation}
\end{widetext}
and the equations of motion corresponding to variations of $\mathbf{A}$, $\mathbf{B}$, $g^{\mu\nu}$ and $\psi$ are given respectively by,
\begin{gather}
d_{\mathbf{A}} \mathbf{B}\wedge dx^\mu = {\rm tr}(\mathbf{B}\wedge \mathbf{B})j^\mu \label{5}\\
\mathbf{F}+\frac{1}{2\kappa}[4\bar{\Lambda}+R(\hat{g})+\kappa T]\mathbf{B} = 0\label{3}\\
R_{\mu\nu}-\frac{1}{4}R(\hat{g})\hat{g}_{\mu\nu} = \kappa(T_{\mu\nu}-\frac{1}{4}T\hat{g}_{\mu\nu})\label{2}\\
\nabla_\mu(\bar{\psi}\gamma^{\mu}) =\bar{\psi}\gamma^{\mu}A_{\mu}
\end{gather}

\noindent where we have defined the fermion current $j^\mu$ as
\be
j^\mu = T^a j^{a\mu}, \qquad j^{a\mu} = g\bar{\psi}_i\gamma^{\mu}(T^a)^i_{\,j}\psi^j.
\ee

\noindent Similarly, we define the axial vector current $j^{5\mu}$ according to
\be
j^{5\mu} = g\bar{\psi}_i\gamma^{\mu}\gamma^{5}\psi^i.
\ee

The equations of motion are supplemented by the canonical anomaly equations of quantum field theory. Namely, the fermion currents $j^\mu$ and $j^{5\mu}$ satisfy
\be
\label{jmuanom}
D_{\mu} j^{\mu}=0 , 
\ee
and
\be
\label{chiralanom}
\nabla_{\mu} j^{\mu5} = \frac{g^2}{4 \pi^2} F \tilde{F} .
\ee
The latter of of these is the standard \emph{chiral anomaly}.  We note that $F \tilde{F}$ is simply the component form of  $\mbox{tr}({\bf F}\wedge {\bf F})$. As such, the chiral anomaly directly relates $\nabla_{\mu} j^{\mu5}$ to the right-hand-side of \eqref{3}, $\frac{1}{2\kappa}[4\bar{\Lambda}+R(\hat{g})+\kappa T]\mathbf{B}$.

The fermions also contribute to the stress-energy tensor of the theory. From the definition \eqref{defTab}, it is given by \cite{Shapiro:2016pfm}
\begin{widetext}
\begin{equation}
T_{\mu\nu} ^{\psi}=-\frac{i}{2}g_{\mu\nu}[\bar{\psi}\gamma^\lambda\nabla_\lambda\psi-\nabla_\lambda\bar{\psi}\gamma^\lambda\psi]+\frac{i}{2}[\bar{\psi}\gamma_{(\mu}\nabla_{\nu)}\psi-\nabla_{(\mu}\bar{\psi}\gamma_{\nu)}\psi]-\frac{1}{2}A_\mu\bar{\psi}\gamma_\nu\psi+m\bar{\psi}\psi g_{\mu\nu}.
\end{equation}
\end{widetext}
This is in addition to other matter contributions to $T_{\mu \nu}$, e.g. from the matter and radiation components of the universe.

This theory exhibits a number of interesting properties which we will now explore.

\subsection{Time-Dependence of Dark Energy}

We define $\Lambda \equiv \tfrac{1}{4}(R + \kappa T) $ as in the theory without fermions. This theory not only allows for dark energy, i.e~a spacetime dependence of the cosmological constant, but instead, in the presence of a net particle number of fermions, \emph{demands} it. We can see this as follows: from the equation of motion of ${\bf A}$, the fermions source a non-zero $d_A {\bf B}$:
\be
{\rm d}_A {\bf B} \wedge {\rm}dx^\mu \propto \bar{\psi} \gamma^\mu \psi.
\ee
 Taking the covariant exterior derivative of \eqref{3},  one finds that ${\rm d}_{\mathbf{A}} {\bf B}$ is a source for dark energy:
\be
{\rm d}  \Lambda \, \wedge B + (\Lambda + \bar\Lambda)  {\rm d}_A {\bf B} = 0.
\label{4}
\ee
Thus, if the fermions have a non-vanishing bilinear $\langle  \bar{\psi} \gamma^\mu \psi \rangle$, the cosmological `constant' is \emph{necessarily} non-constant. Moreover, the spacetime dependence is proportional not to just the gravitating cosmological constant $\Lambda$, but the sum $\Lambda + \bar\Lambda$.

\subsection{Torsion and Non-Conservation of the Stress-Energy Tensor}

It is well known that fermion fields source a torsion contribution to the gravitational connection. More precisely, the connection is given by \cite{freedman2012supergravity}
\begin{equation}
\omega_{\mu\nu\rho}=\omega^{(e)}_{\mu\nu\rho}+K_{\mu\nu\rho},
\end{equation}
where $\omega_{\mu\nu\rho} = \omega_{\mu ab}e^a_{\ \nu} e^b_{\ \rho}$ is the spin connection written in a coordinate basis, ${\omega^{(e)}_{\mu\nu\rho} = \omega^{(e)}_{\mu ab}e^a_{\ \nu} e^b_{\ \rho}}$ is the unique, metric compatible, torsion-free spin connection, and is the $K_{\mu\nu\rho}$ \emph{contorsion} tensor
\begin{equation}
K_{\mu\nu\rho}=-\frac{\kappa}{4}\bar{\psi}\gamma_{\rho\mu\nu}\psi, \qquad \gamma_{\rho\mu\nu}=\left\{\gamma_\rho,\tfrac{1}{2}[\gamma_\mu,\gamma_\nu]\right\},
\end{equation}
defined as the difference between the unique torsion-free Levi-Civita connection and a given torsionful connection that is compatible with the same metric.

This in itself can play an interesting role in cosmology; for example, the torsion induces a 4-fermion interaction, which can trigger a \emph{condensation} of fermions akin to superconductivity \cite{PhysRev.108.1175}. This phenomenon has been proposed as model of dark energy in \cite{Alexander:2009uu}, and a similar condensation as dark matter in \cite{Alexander:2018fjp}.

A familiar feature of torsionful theories is the non-conservation of the stress-tensor. We will now show that in the present model, when the fermions have a non-vanishing bilinear $\langle  \bar{\psi} \gamma^\mu \psi \rangle \neq 0$ inducing a varying effective cosmological constant, the stress tensor is similarly not conserved. 

Taking the covariant derivative of (\ref{2}) with respect to the torsionless connection,
\begin{equation}
\nabla^\mu(R_{\mu\nu}-\frac{1}{4}R g_{\mu\nu})= \nabla^\mu(T_{\mu\nu} - \frac{1}{4}T g_{\mu\nu}),
\end{equation}
and making use of the contracted Bianchi identity $\nabla_\mu R^\mu{}_\nu = \tfrac{1}{2}\partial_\nu R$, one finds, 
\begin{equation}
\nabla^{\mu}T_{\mu\nu}=\frac{1}{4\kappa}\partial_{\nu}\Lambda \label{7}.
\end{equation}
Thus the variation of the cosmological constant, e.g. the time-dependence of dark energy, sources a non-conservation of the stress energy tensor. 

Evidently, the degree to which energy and momentum conservation can be violated in this theory is tied to the variation of the effective cosmological constant. Since we expect only small variations in the cosmological constant are consistent with observed cosmological dynamics, it is natural to assume that any violation of the conservation of the stress tensor is correspondingly small.

\section{Dark Energy as Chiral Asymmetry}

\label{DEasCA}

A striking implication of the matter-coupled theory is a connection between the cosmological constant and the chiral anomaly. If we define $\Lambda$ as the cosmological constant via $R(\hat{g})+\kappa T = 4\Lambda$, inserting equation \eqref{3} into the chiral anomaly \eqref{chiralanom} gives,
\be
\label{eq:chiralanomLambda}
\nabla_{\mu} j^{\mu 5} = \frac{2 g^2}{\pi^2 } (\bar{\Lambda} + \Lambda)^2 .
\ee
We note that, in the solution constructed here, the above implies an equal production of particles and anti-particles. This particle production is thus sourced by the full cosmological constant; both the quantum piece $\bar{\Lambda}$ and the classical piece $\Lambda$ contribute to the breaking of chiral symmetry.  However, by construction, \emph{only the classical piece gravitates}, as in the theory without fermions, equation \eqref{RTLambda}, and observations demand that it must be small. Thus the quantum CC problem must be solved in some other way than the traditional tuning of $\Lambda$ to cancel the quantum contributions.

In what follows we can see this in an explicit solution to the full set of equations of motion. We will see that the quantum contributions can be cancelled via a counter-term contained completely within the BF-sector.

To simplify the analysis, in this section we specialize to an Abelian group,  $G = U(1)$, and assume no direct couplings to other fields, such that constraints on dark U(1)'s, see e.g.~\cite{Knapen:2017xzo,Evans:2017kti,Hewett:2012ns,Cirelli:2016rnw,Dutra:2018gmv} do not apply. We consider an FRW space, with metric  
\be
{\rm d}s^2\equiv g_{\mu \nu} {\rm d}x^\mu {\rm d}x^\nu = - {\rm dt}^2 + a(t)^2 \delta_{ij}{\rm d}x^i{\rm d}x^j   .
\ee
We take an ansatz for the $\mathbf{B}$-field as,
\be
{\bf B} = \frac{1}{3}\sum_{i=1} ^3 {\bf B}^i \label{B-ansatz 1},
\ee
with
\be
\small{
{\bf B}^i = \frac{1}{\sqrt{2}}\left(a(t)^3 f(t) {\rm d}t \wedge {\rm d}x^i + \frac{1}{2f(t)}\epsilon^i{}_{\! jk}{\rm d}x^j \wedge {\rm d}x^k\right). \label{B-ansatz 2}
}
\ee
We will see below that the function $f(t)$ encodes the time-dependence of dark energy. Note that \eqref{B-ansatz 2} implies that
\be
\mathbf{B}^i\wedge\mathbf{B}^j = \delta^{ij}\left(a(t)^3 dt\wedge dx \wedge dy \wedge dz\right),
\ee
so that we have 
\be
\mathbf{B}\wedge \mathbf{B} = a(t)^3 dt\wedge dx \wedge dy \wedge dz = \sqrt{g}\,d^4 x.
\ee

We make an ansatz for the fermion fields that there is a chiral asymmetry but no net particle number, consistent with the anomaly equations \eqref{jmuanom} and \eqref{chiralanom}. We assume that the fermion fields correspond to currents given by
\bea
 j^0=0  \;\; &&,\;\; j^{i} = j^{i}(t)    \\
 j^{05} = j^{05}(t)  \;\; && , \;\;  j^{i5}=0 .
\eea
Physically, this corresponds a system with a chiral asymmetry but no net particle number. Recall that the currents can be expanded as left- and right- handed currents as,
\be
j^0 = n_L + n_R \;\; ,\;\; j^i = j^i _ L -  j^i _R ,
\ee
for the vector current, while for the axial current
\be
j^{05} = n_L - n_R \;\; ,\;\;j^i = j^i _ L + j^i _R .
\ee
We consider a configuration with $n_R = - n_L = \hat{n}$ and $ j^i _ L = - j^i _R $. In this case the only non-vanishing current components are $j^i$ and $j^{05} $. We can easily check that this solves vector current conservation:  $\partial_i (a^3 j^i)=0$ because $j^i$ is space-independent and $\partial_0 (a^3 j^0) =0$ because $j^0=0$.  There are more interesting consequences of the chiral anomaly.

We can now proceed to solve the equations of motion. We define the physical cosmological constant $\Lambda$ as in the fermion free case,
\be
\Lambda \equiv \tfrac{1}{4}(R + \kappa T) .
\ee
The dynamics of $\Lambda$ are governed by the field content. Substituting $\mathbf{B}$ into (\ref{5}), we find
\begin{equation}
j^0=0, \quad j^i=\frac{1}{3\sqrt{2}}\partial_0f^{-1}(t) a^{-3} .
\end{equation}
Inserting this into (\ref{4}), we find
\begin{equation}
\partial_0 \log ( \Lambda + \bar\Lambda)=\partial_0 \log f(t),
\end{equation}
which has solution
\begin{equation}
\Lambda(t)  =\lambda_0 f(t) -\bar\Lambda ,
\label{6}
\end{equation}
where $\lambda_0$ is an integration constant. 

Now that we have established $\Lambda + \bar{\Lambda} = f(t)$, we can solve also for the gauge field $\bf{A}$. Returning to the ansatz \eqref{B-ansatz 1}, \eqref{B-ansatz 2} for the $\bf{B}$ field, and making use of the equation of motion \eqref{3} relating $\bf{B}$ and $\bf{F}$, we find an explicit expression for the field strength two-form
\be
\mathbf{F} = \frac{1}{3}\sum\limits_{i = 1}^{3}\mathbf{F}^i ,
\ee
where
\be
\small{
\mathbf{F}^i = \frac{\lambda_0\sqrt{2}}{\kappa}\left(a(t)^3f(t)^2 dx^i\wedge dt - \frac{1}{2}\epsilon^i{}_{jk}dx^j \wedge dx^k \right).
}
\ee
Noting that for an Abelian gauge group $\mathbf{F} = d \mathbf{A}$, it is straightforward to check that, up to gauge transformations $\mathbf{A} \to \mathbf{A} + d\varphi$, we have
\be 
{\bf A} = \frac{1}{3}\sum_{i=1} ^3 {\bf A}^i  \,,
\ee
with
\be
{\bf A}^i = \frac{\lambda_0\sqrt{2}}{\kappa}\left(a(t)^3f(t)^2 x^i dt - \frac{1}{2}\epsilon^i{}_{jk}x^j dx^k \right).
\ee
This completes the solution to the field content, up to integration constants and the unspecified function $f(t)$.

The chiral anomaly takes a simple form for the above ansatz for the fields. It reads,
\be
\partial_t(n_L-n_R)+3H(n_L-n_R)=\frac{2 g^2}{\pi^2}(\Lambda + \bar\Lambda)^2.
\ee
As anticipated in equation \eqref{eq:chiralanomLambda}, this exhibits a re-emergence of the quantum CC problem: if $\bar{\Lambda}$ is large, this leads to a large fermion production, and the universe will rapidly become dominated by these fermions.

We emphasize that the canonical way to cancel the large quantum contribution to the cosmological constant in the Einstein equation is to fine-tune the bare value of the CC. In the present scenario, the quantum cosmological constant problem has shifted into the chiral anomaly, which is sourced by $ \textrm{tr}(\mathbf{F} \wedge \mathbf{F})$. Moreover, it cannot be cancelled by the classical contribution $\Lambda$, which is distinct from the bare $\bar\Lambda$, since observations demand the physical cosmological constant, $\Lambda$, be small. This demands an entirely new solution to the quantum CC problem.

Interestingly, this cosmological constant problem can be cancelled entirely within the BF sector, via a shift in ${\bf F}$. We can see this as follows. Consider the action with a quantum CC $\bar{\Lambda}$: 
\be
	S = \int \textrm{tr}\left(\mathbf{B}\wedge \mathbf{F} + \bar{\Lambda}\mathbf{B} \wedge \mathbf{B}\right) .
\ee
When $B$ is closed, we can perform a shift ${\bf F}\rightarrow{\bf F}- \Lambda_{\rm c.t.} {\bf B}$, where $c.t.$ stands for ``counter-term'', and the action reads,
\be
S = \int \textrm{tr}\left( \mathbf{B}\wedge \mathbf{F} + (\bar{\Lambda} - \Lambda_{\rm c.t. })\mathbf{B} \wedge \mathbf{B}\right) .
\ee
 This can alternatively be done by introducing a second two form, ${\bf \hat{F}}$, coupled to be ${\bf B}$ via $\int {\bf B} \wedge {\bf \hat{F}}$, and which takes on a background solution $\Lambda_{\rm c.t. } {\bf B}$. In both cases, the variation with respect to $\mathbf{B}$ yields,
\be
\mathbf{F}  =  \left(\Lambda_{\rm c.t.}- \bar{\Lambda}\right)\mathbf{B}.
\ee
This allows for a cancellation of the $\bar{\Lambda} \mathbf{B}$ contribution to $\mathbf{F}$, and hence to the chiral anomaly, via a cancellation of $\bar{\Lambda} \mathbf{B}$ with $\hat{\mathbf{F}}$. An interpretation of this is follows: the above corresponds to a renormalization of the \emph{instanton number} of the BF-sector.

With this in mind, we consider the possibility that the quantum CC problem has been resolved, e.g., by cancelling the quantum contributions with $\Lambda_{\rm c.t.}$. We leave open the possibility that it could alternatively be resolved by an unrelated and independent mechanism for addressing quantum aspects of the CC problem. In any case, this would leave only $\Lambda$ in the chiral anomaly:
\be
\label{chiralAnomLambda}
\partial_t(n_L-n_R)+3H(n_L-n_R)=\frac{2 g^2}{\pi^2}\Lambda ^2 .
\ee
In this case, the global U(1) chiral symmetry is restored in the limit $\Lambda\rightarrow0$.  This implies the resolution to the \emph{new} CC problem, namely the smallness of the observed value of the cosmological constant, is that it is \emph{technically natural} \cite{tHooft:1979rat}, since there is an enhanced symmetry when $\Lambda=0$. 

Technical naturalness arguments for the CC have been made before, e.g. \cite{Aghababaie:2003wz,Burgess:2011va}. In the present context, this arises as a remarkable connection between the physics of the deep IR, namely the curvature of spacetime, and physics of the UV, namely vacuum production of particles. This provides a quantum solution to an {\it a priori}  classical CC problem.

\section{Particle Production from $\Lambda$CDM}

\label{sec:presentday}

The theory we have constructed here connects the vacuum production of fermions to the dark energy content of the universe. An important consistency check is that the production due to the observed cosmological constant does not substantially alter the content or evolution of the universe. To this end, here we will consider the production of particles in our universe given the $\Lambda$CDM parameters, in particular $\Lambda$, as experimentally measured by the {\it Planck} collaboration \cite{aghanim2018planck}.

For simplicity, we take an Abelian group $G=U(1)$, as in the previous section. We fix $\Lambda_{c.t.}=\bar{\Lambda}$ and define,
\be
\Lambda = \Omega_\Lambda H_0^2 a(t)^{-3(1+w_{\rm DE})},
\ee
where $\Omega_\Lambda H_0 ^2$ is the present day cosmological constant (we normalize $a(t_0)=1$), and we define $w_{\rm DE}$ as the equation of state of dark energy. This is equivalent to choosing the function $f(t)$ as,
\be
f(t)=a(t)^{-3(1+w_{\rm DE})} ,
\ee
and similarly $\lambda_0=\Omega_\Lambda H_0^2$. Also recall that in the current setup, the chiral asymmetry corresponds to an equal production of particle and antiparticles, i.e.
\be
n_{tot}=n + \bar{n} = n_R-n_L,
\ee
where we define $n_{tot}$ as the total number of particles and antiparticles. Provided that the number density and gauge coupling are small, the number density is unaffected by annihilations, similar to WIMP cold dark matter.

The present day total number density is given by the integral expression,
\bea
n_{tot}&=&\frac{2 g^2 \Omega_\Lambda^2 H_0^4}{ \pi^2 }\int_0^{t_0} a(t)^{-3(1+2w_{\rm DE})} {\rm d}t.
\eea
This can expressed as an integral over redshift space as,
\be
n_{tot}=\frac{2 g^2 \Omega_\Lambda^2 H_0^4}{ \pi^2 }\int _\infty ^0 \frac{ (1+z)^{2+6w_{\rm DE}}}{H(z)}dz,
\end{equation}
where $H(z)$ is the Hubble parameter. To approximate this, we split the integration into three periods:  dark energy domination ($z<1$),  matter domination ($1 < z < 3000$), and radiation domination ($z> 3000$). This gives,
\bea
n_{tot} \simeq && - \frac{2 g^2 \Omega_\Lambda^2 H_0^4}{ \pi^2 }\Bigg( \int _0^1 \frac{(1+z)^{2+6w_{\rm DE}}}{H_0\sqrt{\Omega_\Lambda}(1+z)^{\frac{3(1+w_{\rm DE})}{2}}}dz \nonumber \\
&&+\int_1^{3000} \frac{(1+z)^{2+6w_{\rm DE}}}{H_0\sqrt{\Omega_M}(1+z)^{\frac{3}{2}}}dz \nonumber\\
&&+\int_{3000}^\infty \frac{(1+z)^{2+6w_{\rm DE}}}{H_0\sqrt{\Omega_R}(1+z)^2} dz \Bigg).
\eea
Following Planck 2018 \cite{aghanim2018planck}, we take $w_{\rm DE}=-0.95$, $\Omega_{\Lambda}=0.6897$,  $\Omega_M=0.3103$, and $\Omega_r=6.9\times 10^{-5}$. We express the number density as a ratio to the entropy density of the universe,  
\begin{equation}
s=\frac{2\pi^2}{45} g_\gamma T_\gamma^3+\frac{2\pi^2}{45}g_\nu T_\nu^3 \simeq 2.3\times 10^{-11}{\rm eV}^3,
\end{equation}
with $g_\gamma=2$ and $g_\nu=2\times7/8$. The resulting chiral asymmetry is,
\begin{equation}
\frac{n_{tot}}{s}=5  \times 10^{-84}g^2.
\end{equation}
Thus we see the small observed cosmological constant corresponds to an extremely small chiral asymmetry for the dark fermion.

This implies that the vacuum particle production induced by the present observed CC is far from observable. It also implies that the technical naturalness of the CC does not simultaneously provide an anthropic preference for a small CC, since large departures in $\Lambda$ would leave $n_{tot}$ still far from observable, let alone inhospitable to life. However, we emphasize that the lack of an anthropic argument has no relevance to the status of the technical naturalness; on the contrary, technical naturalness is appealing as an alternative to anthropics.

\section{Inflationary Production of Dark Matter}

\label{sec:inflation}

We have seen the present acceleration  of the universe generates a negligible particle number. However, the far past of the universe may have undergone an accelerating phase with a far greater vacuum energy. This is the central hypothesis of inflationary cosmology \cite{Guth:1980zm}, which posits a phase of quasi-de Sitter expansion.

The inflationary universe can be realized in the present model through the time-dependence of $\Lambda$, encoded in \eqref{6} by the functional freedom to choose $f(t)$. Inflation driven purely by $\Lambda$, with no other field content, corresponds to a slowly-varying $f(t)$, and inflation ends at some time at which $f(t)$ undergoes a steep drop in value. More generally, $\Lambda(t)$ may be sub-dominant but non-negligible to the inflationary energy density, with inflation proceeding along conventional lines, e.g. via a scalar field $\varphi$ with a potential $V(\varphi)$.

An interesting cosmological possibility is that the origins of dark matter could lie in the physics of inflation. This idea has been widely explored, e.g. \cite{Alexander:2018fjp,Tenkanen:2019aij,Herring:2019hbe,Ema:2019yrd,Chung:2011ck,Graham:2015rva,Li:2020xwr} Here we explore this possibility in the present context, with particle production induced by the chiral anomaly \eqref{chiralAnomLambda}.

We take $\Lambda \simeq H^2$ a constant, where $H$ is the Hubble scale during inflation. We write the chiral anomaly, i.e. the equation of motion for $n_{tot}$, using as time coordinate the number of e-folds of expansion, defined as ${\rm d}N=H {\rm d}t$. We find
\be
\frac{1}{a^3} \frac{{\rm d}}{{\rm d}N} a^3 n_{tot} \simeq \frac{2 g^2}{ \pi^2} H^3 .
\ee
The solution to this is given by,
\be
n_{tot}(N) = \frac{2 g^2}{3 \pi^2} H^3 + c H^3 e^{-3 N},
\ee
where $c$ is an integration constant. For a large number of e-folds of inflation, the second term is negligible. Thus we can approximate,
\be
 n_{tot}(t_i) \simeq \frac{2 g^2}{3 \pi^2} H^3 ,
\ee
where $t_i$ is the end of inflation. This constitutes the total number of particles at the end of inflation. The number density at subsequent times, provided the gauge coupling and density are too low to allow significant particle-antiparticle annihilation, and neglecting the small production considered in the previous subsection, simply scales as matter,
\be
\label{ntott}
n_{tot}(t) = \frac{2 g^2}{3 \pi^2 a(t)^3} H_i ^3.
\ee
where $H_i$ refers to the Hubble scale during inflation. This follows directly from solving \eqref{6}, with $\Lambda \simeq 0$ in the post-inflationary universe. We now seek to understand the circumstances under which this can constitute the dark matter in our universe.

An Abelian group will not suffice. For massless particles and an Abelian gauge group, the subsequent cosmological dynamics are boring: the energy density redshifts as radiation, and hence the fermions constitute a form of dark radiation, that can can never come to dominate the universe. One could of course ameliorate this by adding additional fields, e.g. a dark Higgs field which generates a mass for the fermions, but will not consider such detailed model building here.

For a non-Abelian group, much more can happen. In particular, this opens the possibility of confinement, such that dark matter today is in the form of dark baryons. This idea has been explored in, e.g., \cite{Antipin:2015xia,Hertzberg:2019bvt}.  With primordial baryons, the cosmological evolution is similar that of the ``superheavy dark matter'' scenario \cite{Li:2020xwr,Kolb:1998ki,Chung:1998zb,Berezinsky:1997hy,Kuzmin:1997jua,Birkel:1998nx,Bhattacharjee:1998qc}: the primordial abundance of dark matter redshifts like matter, such that the number of particles per Hubble volume grows with the scale factor during radiation domination, and can provide all the dark matter in the universe today.  The superheavy DM in the present case is dark baryons; or more precisely, an equal and opposite amount of dark baryons and dark anti-baryons.

We can see this in a concrete example. We consider the group $SU(3)$, and two flavours of Dirac fermion, which play the role of the up and down quark of the dark QCD theory. There are 3 particles per baryon, and hence the total number density of dark baryons is $n_{tot}/3$.  The mass of the dark baryons is predominantly determined by the binding energy, which scales with the strong coupling of the theory. We denote the dark sector strong coupling scale as $\Lambda_{D}$. Absent an experimental measurement that anchors the coupling at some energy scale, the strong coupling scale is a free parameter. The energy in the dark baryons is then given by,
\be
\rho_{\rm DM} \simeq \frac{1}{3}\Lambda_D n_{tot}(t_c),
\ee
where $t_c$ is the time at which the fermions confine, related to the fermion density produced by inflation by equation \eqref{ntott}.

One can easily compute the requisite $\Lambda_{D}$ for the fermions produced during inflation to be all the dark matter today.  As in canonical inflationary production of dark matter, there is very little impact of baryon-antibaryon annihilation on the number density, since the primordial density is extremely low relative to standard thermal history for dark matter, and the system is far from equilibrium. Using the results of e.g. \cite{Li:2020xwr}, we find the present abundance,  
\bea
\Omega_{\rm DM} h^2 &\simeq& \frac{8 \pi}{3} \Omega_{rad}h^2 \, \frac{\Lambda_D n_{tot}(t_i)}{m_{pl}^2 H^2} \left(\frac{T_{ re}}{T_0} \right) \nonumber \\  &\simeq& 10^9 \Lambda_D \frac{H T_{ re}}{M_{pl}^2 {\rm GeV}},
\eea
where $T_{re}$ is the reheating temperature and $T_0$ is the present CMB temperature. Assuming instant reheating, $H^2 \simeq T_{re}^4/(m_{pl})^2$, this is
\be
\label{relicdens}
\Omega_{\rm DM}h^2 \simeq 10^9 \frac{\Lambda_D}{\rm GeV} \left( \frac{H}{M_{pl}}\right)^{3/2}.
\ee
Demanding this be the observed abundance, 0.12, relates the strong coupling scale $\Lambda_D$ to the Hubble scale of inflation, as
\be
\frac{\Lambda_D}{\rm GeV} = \left( \frac{5.92 \times 10^{11} {\rm GeV}}{H} \right)^{\frac{3}{2}}.
\ee
For $\Lambda_D$ and $H$ satisfying the above, the chiral anomaly during inflation is sufficient to produce the observed dark matter abundance. As a simple example, for $\Lambda_D \sim H$, one finds $H\gtrsim 10^{7}$ GeV gives the correct relic density of dark matter, in the form of dark baryons of mass $\Lambda_D \sim 10^7 {\rm GeV}$, constituting a realization of superheavy dark matter.

We postpone an in depth analysis of this dark matter scenario for future work. For example, a precise calculation of the number density including the non-adiabatic evolution of the vacuum state at the end of inflation \cite{Kolb:1998ki,Chung:1998zb}, the impact of known constraints on dark baryon dark matter \cite{Antipin:2015xia}, and the impact of CMB isocurvature constraints \cite{Akrami:2018odb}. 
One could also consider dark mesons in addition to or in place of dark baryons. For the moment we conclude that the matter coupled BF-gravity provides a connection between inflation and dark matter production, which may be able to produce the entire observed abundance of dark matter.

\section{Discussion}

\label{sec:conclusion}

The cosmological constant has both inspired and confounded physicists for decades. One approach to this problem is to modify the Einstein-Hilbert action in such a way that the cosmological constant, or at least the piece of it which gravitates, is not renormalized. A specific proposal along these lines (\cite{Alexander:2018tyf}) is to couple gravity to a topological field theory, and have the fields of the latter theory absorb all quantum corrections to the vacuum energy.

In this work we have extended \cite{Alexander:2018tyf} to include matter charged under the BF gauge group. We have constructed FRW solutions satisfying all the equations of motion, including one with a striking implication: the chiral anomaly. We find, eq.~\eqref{chiralanom}, that the $U(1)_A$ axial symmetry of charged fermions is broken by a non-zero cosmological constant. In the limit of vanishing cosmological constant, the symmetry is restored. This makes a small cosmological constant \emph{technically natural} \cite{tHooft:1979rat}.

A consequence of this chiral anomaly is that any non-zero cosmological constant will source the production of particles out of the vacuum. Given the observed $\Lambda$CDM parameters, this is a miniscule effect. However, in the early universe, when the vacuum energy of the universe may have been much larger (e.g.~ during inflation), the effect can be drastic. In particular, the production of particles can explain the entire observed dark matter abundance.

There are many directions for future work. The dynamics of cosmological perturbations and the formation of structure remain to be studied in the scenario of \cite{Alexander:2018tyf}. In particular, applied to the early universe, any modification to the dynamics of cosmological perturbations may be relevant to inflationary model building and the predictions of inflation.  It will also be interesting to consider the realization of alternatives to inflation, such as non-singular bouncing cosmology (see e.g. \cite{Brandenberger:2016vhg,Cai:2012va,Cai:2013kja,Alexander:2014eva}). Finally, a striking implication of the setup considered here is that a cosmological \emph{constant} is in general not possible, and instead, by equation \eqref{4}, $\Lambda$ has a spacetime dependence dictated by the fermion vector current. This is especially interesting considering the renewed debate as to the existence of exact de Sitter space in quantum gravity \cite{Obied:2018sgi}, see also e.g. \cite{Kallosh:2019axr}, and slightly older works such as e.g.~\cite{Dasgupta:2014pma}. We leave these interesting possibilities to future work.

 \section*{Acknowledgements}

The authors thank Jatan Buch, Shing Chau (John) Leung, Sylvester James Gates Jr., Rachel Houtz, Leah Jenks, and David Ramirez, for helpful discussions.

\bibliographystyle{JHEP}
\bibliography{BF-fermions-refs}

\providecommand{\href}[2]{#2}\begingroup\raggedright\begin{thebibliography}{10}

\bibitem{Zeldovich:1967gd}
Y.~B. Zeldovich, \emph{{Cosmological Constant and Elementary Particles}},
  {\emph{JETP Lett.} {\bfseries 6} (1967) 316}.

\bibitem{weinberg1989cosmological}
S.~Weinberg, \emph{The cosmological constant problem}, {\emph{Reviews of modern
  physics} {\bfseries 61} (1989) 1}.

\bibitem{padilla2015lectures}
A.~Padilla, \emph{Lectures on the cosmological constant problem}, {\emph{arXiv
  preprint arXiv:1502.05296} (2015) }.

\bibitem{Bousso2007TheCC}
R.~Bousso, \emph{The cosmological constant}, {\emph{General Relativity and
  Gravitation} {\bfseries 40} (2007) 607}.

\bibitem{perlmutter1999measurements}
S.~Perlmutter, G.~Aldering, G.~Goldhaber, R.~Knop, P.~Nugent, P.~Castro et~al.,
  \emph{Measurements of $\omega$ and $\lambda$ from 42 high-redshift
  supernovae}, {\emph{The Astrophysical Journal} {\bfseries 517} (1999) 565}.

\bibitem{riess1998observational}
A.~G. Riess, A.~V. Filippenko, P.~Challis, A.~Clocchiatti, A.~Diercks, P.~M.
  Garnavich et~al., \emph{Observational evidence from supernovae for an
  accelerating universe and a cosmological constant}, {\emph{The Astronomical
  Journal} {\bfseries 116} (1998) 1009}.

\bibitem{2013PhR...530...87W}
D.~H. {Weinberg}, M.~J. {Mortonson}, D.~J. {Eisenstein}, C.~{Hirata}, A.~G.
  {Riess} and E.~{Rozo}, \emph{{Observational probes of cosmic acceleration}},
  \href{https://doi.org/10.1016/j.physrep.2013.05.001}{\emph{Phys. Rep.}
  {\bfseries 530} (2013) 87} [\href{https://arxiv.org/abs/1201.2434}{{\ttfamily
  1201.2434}}].

\bibitem{aghanim2018planck}
N.~Aghanim, Y.~Akrami, M.~Ashdown, J.~Aumont, C.~Baccigalupi, M.~Ballardini
  et~al., \emph{Planck 2018 results. vi. cosmological parameters}, {\emph{arXiv
  preprint arXiv:1807.06209} (2018) }.

\bibitem{Tseytlin:1990hn}
A.~A. Tseytlin, \emph{{Duality symmetric string theory and the cosmological
  constant problem}},
  \href{https://doi.org/10.1103/PhysRevLett.66.545}{\emph{Phys. Rev. Lett.}
  {\bfseries 66} (1991) 545}.

\bibitem{Gabadadze:2014rwa}
G.~Gabadadze, \emph{{The Big Constant Out, The Small Constant In}},
  \href{https://doi.org/10.1016/j.physletb.2014.10.064}{\emph{Phys. Lett.}
  {\bfseries B739} (2014) 263}
  [\href{https://arxiv.org/abs/1406.6701}{{\ttfamily 1406.6701}}].

\bibitem{Gabadadze:2016avp}
G.~Gabadadze and S.~Yu, \emph{{Dark energy and inflation with a volume
  normalized action}},
  \href{https://doi.org/10.1103/PhysRevD.98.024047}{\emph{Phys. Rev.}
  {\bfseries D98} (2018) 024047}
  [\href{https://arxiv.org/abs/1611.05833}{{\ttfamily 1611.05833}}].

\bibitem{Gabadadze:2015goa}
G.~Gabadadze and S.~Yu, \emph{{Metamorphosis of the Cosmological Constant and
  5D Origin of the Fiducial Metric}},
  \href{https://doi.org/10.1103/PhysRevD.94.104059}{\emph{Phys. Rev.}
  {\bfseries D94} (2016) 104059}
  [\href{https://arxiv.org/abs/1510.07943}{{\ttfamily 1510.07943}}].

\bibitem{Kaloper:2013zca}
N.~Kaloper and A.~Padilla, \emph{{Sequestering the Standard Model Vacuum
  Energy}}, \href{https://doi.org/10.1103/PhysRevLett.112.091304}{\emph{Phys.
  Rev. Lett.} {\bfseries 112} (2014) 091304}
  [\href{https://arxiv.org/abs/1309.6562}{{\ttfamily 1309.6562}}].

\bibitem{Kaloper:2014dqa}
N.~Kaloper and A.~Padilla, \emph{{Vacuum Energy Sequestering: The Framework and
  Its Cosmological Consequences}},
  \href{https://doi.org/10.1103/PhysRevD.90.084023,
  10.1103/PhysRevD.90.109901}{\emph{Phys. Rev.} {\bfseries D90} (2014) 084023}
  [\href{https://arxiv.org/abs/1406.0711}{{\ttfamily 1406.0711}}].

\bibitem{Kaloper:2016jsd}
N.~Kaloper and A.~Padilla, \emph{{Vacuum Energy Sequestering and Graviton
  Loops}}, \href{https://doi.org/10.1103/PhysRevLett.118.061303}{\emph{Phys.
  Rev. Lett.} {\bfseries 118} (2017) 061303}
  [\href{https://arxiv.org/abs/1606.04958}{{\ttfamily 1606.04958}}].

\bibitem{Alexander:2018tyf}
S.~Alexander and R.~Carballo-Rubio, \emph{{Topological Features of the Quantum
  Vacuum}},  \href{https://arxiv.org/abs/1810.02159}{{\ttfamily 1810.02159}}.

\bibitem{Aharony:2008ug}
O.~Aharony, O.~Bergman, D.~L. Jafferis and J.~Maldacena, \emph{{N=6
  superconformal Chern-Simons-matter theories, M2-branes and their gravity
  duals}}, \href{https://doi.org/10.1088/1126-6708/2008/10/091}{\emph{JHEP}
  {\bfseries 10} (2008) 091} [\href{https://arxiv.org/abs/0806.1218}{{\ttfamily
  0806.1218}}].

\bibitem{Gates:1991qn}
S.~J. Gates, Jr. and H.~Nishino, \emph{{Remarks on the N=2 supersymmetric
  Chern-Simons theories}},
  \href{https://doi.org/10.1016/0370-2693(92)90277-B}{\emph{Phys. Lett.}
  {\bfseries B281} (1992) 72}.

\bibitem{Nishino:1991sr}
H.~Nishino and S.~J. Gates, Jr., \emph{{Chern-Simons theories with
  supersymmetries in three-dimensions}},
  \href{https://doi.org/10.1142/S0217751X93001363}{\emph{Int. J. Mod. Phys.}
  {\bfseries A8} (1993) 3371}.

\bibitem{Nishino:1996xb}
H.~Nishino and S.~J. Gates, Jr., \emph{{Aleph0 extended supergravity and
  Chern-Simons theories}},
  \href{https://doi.org/10.1016/S0550-3213(96)00476-2}{\emph{Nucl. Phys.}
  {\bfseries B480} (1996) 573}
  [\href{https://arxiv.org/abs/hep-th/9606090}{{\ttfamily hep-th/9606090}}].

\bibitem{tHooft:1979rat}
G.~'t~Hooft, \emph{{Naturalness, chiral symmetry, and spontaneous chiral
  symmetry breaking}},
  \href{https://doi.org/10.1007/978-1-4684-7571-5_9}{\emph{NATO Sci. Ser. B}
  {\bfseries 59} (1980) 135}.

\bibitem{Guth:1980zm}
A.~H. Guth, \emph{{The Inflationary Universe: A Possible Solution to the
  Horizon and Flatness Problems}},
  \href{https://doi.org/10.1103/PhysRevD.23.347}{\emph{Phys. Rev.} {\bfseries
  D23} (1981) 347}.

\bibitem{Antipin:2015xia}
O.~Antipin, M.~Redi, A.~Strumia and E.~Vigiani, \emph{{Accidental Composite
  Dark Matter}}, \href{https://doi.org/10.1007/JHEP07(2015)039}{\emph{JHEP}
  {\bfseries 07} (2015) 039}
  [\href{https://arxiv.org/abs/1503.08749}{{\ttfamily 1503.08749}}].

\bibitem{Hertzberg:2019bvt}
M.~P. Hertzberg and M.~Sandora, \emph{{Dark Matter and Naturalness}},
  \href{https://doi.org/10.1007/JHEP12(2019)037}{\emph{JHEP} {\bfseries 12}
  (2019) 037} [\href{https://arxiv.org/abs/1908.09841}{{\ttfamily
  1908.09841}}].

\bibitem{cattaneo1995topological}
A.~S. Cattaneo, P.~Cotta-Ramusino, J.~Fr{\"o}hlich and M.~Martellini,
  \emph{Topological bf theories in 3 and 4 dimensions}, {\emph{Journal of
  Mathematical Physics} {\bfseries 36} (1995) 6137}.

\bibitem{Brooks:1994nn}
R.~Brooks and S.~J. Gates, Jr., \emph{{Extended supersymmetry and superBF gauge
  theories}}, \href{https://doi.org/10.1016/0550-3213(94)90600-9}{\emph{Nucl.
  Phys.} {\bfseries B432} (1994) 205}
  [\href{https://arxiv.org/abs/hep-th/9407147}{{\ttfamily hep-th/9407147}}].

\bibitem{10.1007/3-540-46552-9_2}
J.~C. Baez, \emph{An introduction to spin foam models of bf theory and quantum
  gravity},  in \emph{Geometry and Quantum Physics} (H.~Gausterer, L.~Pittner
  and H.~Grosse, eds.), (Berlin, Heidelberg), pp.~25--93, Springer Berlin
  Heidelberg, 2000.

\bibitem{doi:10.1063/1.531238}
A.~S. Cattaneo, P.~Cotta‐Ramusino, J.~Fröhlich and M.~Martellini,
  \emph{Topological bf theories in 3 and 4 dimensions},
  \href{https://doi.org/10.1063/1.531238}{\emph{Journal of Mathematical
  Physics} {\bfseries 36} (1995) 6137}
  [\href{https://arxiv.org/abs/https://doi.org/10.1063/1.531238}{{\ttfamily
  https://doi.org/10.1063/1.531238}}].

\bibitem{Shapiro:2016pfm}
I.~L. Shapiro, \emph{{Covariant derivative of fermions and all that}},
  \href{https://arxiv.org/abs/1611.02263}{{\ttfamily 1611.02263}}.

\bibitem{freedman2012supergravity}
D.~Z. Freedman and A.~Van~Proeyen, \emph{Supergravity}. Cambridge university
  press, 2012.

\bibitem{PhysRev.108.1175}
J.~Bardeen, L.~N. Cooper and J.~R. Schrieffer, \emph{Theory of
  superconductivity},
  \href{https://doi.org/10.1103/PhysRev.108.1175}{\emph{Phys. Rev.} {\bfseries
  108} (1957) 1175}.

\bibitem{Alexander:2009uu}
S.~Alexander, T.~Biswas and G.~Calcagni, \emph{{Cosmological
  Bardeen-Cooper-Schrieffer condensate as dark energy}},
  \href{https://doi.org/10.1103/PhysRevD.81.069902,
  10.1103/PhysRevD.81.043511}{\emph{Phys. Rev.} {\bfseries D81} (2010) 043511}
  [\href{https://arxiv.org/abs/0906.5161}{{\ttfamily 0906.5161}}].

\bibitem{Alexander:2018fjp}
S.~Alexander, E.~McDonough and D.~N. Spergel, \emph{{Chiral Gravitational Waves
  and Baryon Superfluid Dark Matter}},
  \href{https://doi.org/10.1088/1475-7516/2018/05/003}{\emph{JCAP} {\bfseries
  1805} (2018) 003} [\href{https://arxiv.org/abs/1801.07255}{{\ttfamily
  1801.07255}}].

\bibitem{Knapen:2017xzo}
S.~Knapen, T.~Lin and K.~M. Zurek, \emph{{Light Dark Matter: Models and
  Constraints}}, \href{https://doi.org/10.1103/PhysRevD.96.115021}{\emph{Phys.
  Rev.} {\bfseries D96} (2017) 115021}
  [\href{https://arxiv.org/abs/1709.07882}{{\ttfamily 1709.07882}}].

\bibitem{Evans:2017kti}
J.~A. Evans, S.~Gori and J.~Shelton, \emph{{Looking for the WIMP Next Door}},
  \href{https://doi.org/10.1007/JHEP02(2018)100}{\emph{JHEP} {\bfseries 02}
  (2018) 100} [\href{https://arxiv.org/abs/1712.03974}{{\ttfamily
  1712.03974}}].

\bibitem{Hewett:2012ns}
\emph{{Fundamental Physics at the Intensity Frontier}}, 2012.
\newblock 10.2172/1042577.

\bibitem{Cirelli:2016rnw}
M.~Cirelli, P.~Panci, K.~Petraki, F.~Sala and M.~Taoso, \emph{{Dark Matter's
  secret liaisons: phenomenology of a dark U(1) sector with bound states}},
  \href{https://doi.org/10.1088/1475-7516/2017/05/036}{\emph{JCAP} {\bfseries
  1705} (2017) 036} [\href{https://arxiv.org/abs/1612.07295}{{\ttfamily
  1612.07295}}].

\bibitem{Dutra:2018gmv}
M.~Dutra, M.~Lindner, S.~Profumo, F.~S. Queiroz, W.~Rodejohann and C.~Siqueira,
  \emph{{MeV Dark Matter Complementarity and the Dark Photon Portal}},
  \href{https://doi.org/10.1088/1475-7516/2018/03/037}{\emph{JCAP} {\bfseries
  1803} (2018) 037} [\href{https://arxiv.org/abs/1801.05447}{{\ttfamily
  1801.05447}}].

\bibitem{Aghababaie:2003wz}
Y.~Aghababaie, C.~P. Burgess, S.~L. Parameswaran and F.~Quevedo, \emph{{Towards
  a naturally small cosmological constant from branes in 6-D supergravity}},
  \href{https://doi.org/10.1016/j.nuclphysb.2003.12.015}{\emph{Nucl. Phys.}
  {\bfseries B680} (2004) 389}
  [\href{https://arxiv.org/abs/hep-th/0304256}{{\ttfamily hep-th/0304256}}].

\bibitem{Burgess:2011va}
C.~P. Burgess and L.~van Nierop, \emph{{Technically Natural Cosmological
  Constant From Supersymmetric 6D Brane Backreaction}},
  \href{https://doi.org/10.1016/j.dark.2012.10.001}{\emph{Phys. Dark Univ.}
  {\bfseries 2} (2013) 1} [\href{https://arxiv.org/abs/1108.0345}{{\ttfamily
  1108.0345}}].

\bibitem{Tenkanen:2019aij}
T.~Tenkanen, \emph{{Dark matter from scalar field fluctuations}},
  \href{https://doi.org/10.1103/PhysRevLett.123.061302}{\emph{Phys. Rev. Lett.}
  {\bfseries 123} (2019) 061302}
  [\href{https://arxiv.org/abs/1905.01214}{{\ttfamily 1905.01214}}].

\bibitem{Herring:2019hbe}
N.~Herring, D.~Boyanovsky and A.~R. Zentner, \emph{{Non-adiabatic cosmological
  production of ultra-light Dark Matter}},
  \href{https://arxiv.org/abs/1912.10859}{{\ttfamily 1912.10859}}.

\bibitem{Ema:2019yrd}
Y.~Ema, K.~Nakayama and Y.~Tang, \emph{{Production of Purely Gravitational Dark
  Matter: The Case of Fermion and Vector Boson}},
  \href{https://doi.org/10.1007/JHEP07(2019)060}{\emph{JHEP} {\bfseries 07}
  (2019) 060} [\href{https://arxiv.org/abs/1903.10973}{{\ttfamily
  1903.10973}}].

\bibitem{Chung:2011ck}
D.~J.~H. Chung, L.~L. Everett, H.~Yoo and P.~Zhou, \emph{{Gravitational Fermion
  Production in Inflationary Cosmology}},
  \href{https://doi.org/10.1016/j.physletb.2012.04.066}{\emph{Phys. Lett.}
  {\bfseries B712} (2012) 147}
  [\href{https://arxiv.org/abs/1109.2524}{{\ttfamily 1109.2524}}].

\bibitem{Graham:2015rva}
P.~W. Graham, J.~Mardon and S.~Rajendran, \emph{{Vector Dark Matter from
  Inflationary Fluctuations}},
  \href{https://doi.org/10.1103/PhysRevD.93.103520}{\emph{Phys. Rev.}
  {\bfseries D93} (2016) 103520}
  [\href{https://arxiv.org/abs/1504.02102}{{\ttfamily 1504.02102}}].

\bibitem{Li:2020xwr}
L.~Li, S.~Lu, Y.~Wang and S.~Zhou, \emph{{Cosmological Signatures of Superheavy
  Dark Matter}},  \href{https://arxiv.org/abs/2002.01131}{{\ttfamily
  2002.01131}}.

\bibitem{Kolb:1998ki}
E.~W. Kolb, D.~J.~H. Chung and A.~Riotto, \emph{{WIMPzillas!}},
  \href{https://doi.org/10.1063/1.59655}{\emph{AIP Conf. Proc.} {\bfseries 484}
  (1999) 91} [\href{https://arxiv.org/abs/hep-ph/9810361}{{\ttfamily
  hep-ph/9810361}}].

\bibitem{Chung:1998zb}
D.~J.~H. Chung, E.~W. Kolb and A.~Riotto, \emph{{Superheavy dark matter}},
  \href{https://doi.org/10.1103/PhysRevD.59.023501}{\emph{Phys. Rev.}
  {\bfseries D59} (1998) 023501}
  [\href{https://arxiv.org/abs/hep-ph/9802238}{{\ttfamily hep-ph/9802238}}].

\bibitem{Berezinsky:1997hy}
V.~Berezinsky, M.~Kachelriess and A.~Vilenkin, \emph{{Ultrahigh-energy cosmic
  rays without GZK cutoff}},
  \href{https://doi.org/10.1103/PhysRevLett.79.4302}{\emph{Phys. Rev. Lett.}
  {\bfseries 79} (1997) 4302}
  [\href{https://arxiv.org/abs/astro-ph/9708217}{{\ttfamily
  astro-ph/9708217}}].

\bibitem{Kuzmin:1997jua}
V.~A. Kuzmin and V.~A. Rubakov, \emph{{Ultrahigh-energy cosmic rays: A Window
  to postinflationary reheating epoch of the universe?}}, {\emph{Phys. Atom.
  Nucl.} {\bfseries 61} (1998) 1028}
  [\href{https://arxiv.org/abs/astro-ph/9709187}{{\ttfamily
  astro-ph/9709187}}].

\bibitem{Birkel:1998nx}
M.~Birkel and S.~Sarkar, \emph{{Extremely high-energy cosmic rays from relic
  particle decays}},
  \href{https://doi.org/10.1016/S0927-6505(98)00028-0}{\emph{Astropart. Phys.}
  {\bfseries 9} (1998) 297}
  [\href{https://arxiv.org/abs/hep-ph/9804285}{{\ttfamily hep-ph/9804285}}].

\bibitem{Bhattacharjee:1998qc}
P.~Bhattacharjee and G.~Sigl, \emph{{Origin and propagation of extremely
  high-energy cosmic rays}},
  \href{https://doi.org/10.1016/S0370-1573(99)00101-5}{\emph{Phys. Rept.}
  {\bfseries 327} (2000) 109}
  [\href{https://arxiv.org/abs/astro-ph/9811011}{{\ttfamily
  astro-ph/9811011}}].

\bibitem{Akrami:2018odb}
{\scshape Planck} collaboration, Y.~Akrami et~al., \emph{{Planck 2018 results.
  X. Constraints on inflation}},
  \href{https://arxiv.org/abs/1807.06211}{{\ttfamily 1807.06211}}.

\bibitem{Brandenberger:2016vhg}
R.~Brandenberger and P.~Peter, \emph{{Bouncing Cosmologies: Progress and
  Problems}}, \href{https://doi.org/10.1007/s10701-016-0057-0}{\emph{Found.
  Phys.} {\bfseries 47} (2017) 797}
  [\href{https://arxiv.org/abs/1603.05834}{{\ttfamily 1603.05834}}].

\bibitem{Cai:2012va}
Y.-F. Cai, D.~A. Easson and R.~Brandenberger, \emph{{Towards a Nonsingular
  Bouncing Cosmology}},
  \href{https://doi.org/10.1088/1475-7516/2012/08/020}{\emph{JCAP} {\bfseries
  1208} (2012) 020} [\href{https://arxiv.org/abs/1206.2382}{{\ttfamily
  1206.2382}}].

\bibitem{Cai:2013kja}
Y.-F. Cai, E.~McDonough, F.~Duplessis and R.~H. Brandenberger, \emph{{Two Field
  Matter Bounce Cosmology}},
  \href{https://doi.org/10.1088/1475-7516/2013/10/024}{\emph{JCAP} {\bfseries
  1310} (2013) 024} [\href{https://arxiv.org/abs/1305.5259}{{\ttfamily
  1305.5259}}].

\bibitem{Alexander:2014eva}
S.~Alexander, C.~Bambi, A.~Marciano and L.~Modesto, \emph{{Fermi-bounce
  Cosmology and scale invariant power-spectrum}},
  \href{https://doi.org/10.1103/PhysRevD.90.123510}{\emph{Phys. Rev.}
  {\bfseries D90} (2014) 123510}
  [\href{https://arxiv.org/abs/1402.5880}{{\ttfamily 1402.5880}}].

\bibitem{Obied:2018sgi}
G.~Obied, H.~Ooguri, L.~Spodyneiko and C.~Vafa, \emph{{De Sitter Space and the
  Swampland}},  \href{https://arxiv.org/abs/1806.08362}{{\ttfamily
  1806.08362}}.

\bibitem{Kallosh:2019axr}
R.~Kallosh, A.~Linde, E.~McDonough and M.~Scalisi, \emph{{dS Vacua and the
  Swampland}}, \href{https://doi.org/10.1007/JHEP03(2019)134}{\emph{JHEP}
  {\bfseries 03} (2019) 134}
  [\href{https://arxiv.org/abs/1901.02022}{{\ttfamily 1901.02022}}].

\bibitem{Dasgupta:2014pma}
K.~Dasgupta, R.~Gwyn, E.~McDonough, M.~Mia and R.~Tatar, \emph{{de Sitter Vacua
  in Type IIB String Theory: Classical Solutions and Quantum Corrections}},
  \href{https://doi.org/10.1007/JHEP07(2014)054}{\emph{JHEP} {\bfseries 07}
  (2014) 054} [\href{https://arxiv.org/abs/1402.5112}{{\ttfamily 1402.5112}}].

\end{thebibliography}\endgroup

\end{document}